\documentclass[12,preprint]{aastex}

\usepackage{graphicx}
\usepackage{rotating}
\usepackage{lscape}

\newcommand{\um}{$\micron$}

\newcommand{\fa}{$F_{6.4}$}

\newcommand{\fb}{$F_{6.6}$}
\newcommand{\fc}{$F_{7.7}$}

\newcommand{\fn}{$F_{19.4}$}
\newcommand{\fe}{$F_{24.2}$}
\newcommand{\ff}{$F_{31.5}$}
\newcommand{\fg}{$F_{37.1}$}
\newcommand{\td}{$T_{Dust}$}

\shorttitle{SOFIA-FORCAST view of W3A}
\shortauthors{Salgado et al.}

\begin{document}

\title{First science results from SOFIA/FORCAST: The mid-infrared view of the compact HII region W3A}

\author{F. Salgado\altaffilmark{1}, O. Bern\'e\altaffilmark{1}, 
J. D. Adams\altaffilmark{2}, T. L. Herter\altaffilmark{2}, G. Gull\altaffilmark{2}, J. Schoenwald\altaffilmark{2},  
L. D. Keller\altaffilmark{3}, J. M. De Buizer\altaffilmark{4},
W. D. Vacca\altaffilmark{4}, E. E. Becklin\altaffilmark{4}, R. Y. Shuping\altaffilmark{4}, A. G. G. M., Tielens\altaffilmark{1}, and H. Zinnecker\altaffilmark{5,6}
}

\altaffiltext{1}{Leiden Observatory, University of Leiden, P. O. Box 9513, 2300 RA Leiden, Netherlands}
\altaffiltext{2}{Astronomy Department, 202 Space Sciences Building, Cornell University, Ithaca, NY 14853-6801, USA}
\altaffiltext{3}{Department of Physics, Ithaca College, Ithaca, NY 14850, USA}
\altaffiltext{4}{SOFIA-USRA, NASA Ames Research Center, MS N211-3, Moffett Field, CA 94035, USA}
\altaffiltext{5}{SOFIA Science Center, NASA Ames Research Center, MS N211-3, Moffett Field, CA 94035, USA}
\altaffiltext{6}{Deutsches SOFIA Institut, Univ. Stuttgart, Germany}

\begin{abstract}
The massive star forming region W3 was observed with the faint object infrared camera for the SOFIA telescope (FORCAST) as part of the
Short Science program. The 6.4, 6.6, 7.7, 19.7, 24.2, 31.5 and 37.1 \um~bandpasses were used to observe the emission of Polycyclic Aromatic Hydrocarbon (PAH) molecules, Very Small Grains and Big Grains. Optical depth and color temperature maps of W3A show that IRS2 has blown a bubble devoid of gas and dust of $\sim$0.05~pc radius. It is embedded in a dusty shell of ionized gas that contributes 40\% of the total 24 \um~emission of W3A. This dust component is mostly heated by far ultraviolet, rather than trapped Ly$\alpha$~photons. This shell is itself surrounded by a thin ($\sim$0.01~pc) photodissociation region where PAHs show intense emission. The infrared spectral energy distribution (SED) of three different zones located at 8, 20 and 25\arcsec~from IRS2, show that the peak of the SED shifts towards longer wavelengths, when moving away from the star. Adopting the stellar radiation field for these three positions,  DUSTEM model fits to these SEDs yield a dust-to-gas mass ratio in the ionized gas similar to that in the diffuse ISM. However, the ratio of the IR-to-UV opacity of the dust in the ionized shell is  increased by a factor $\simeq$3~compared to the diffuse ISM.
\end{abstract}

\section{Introduction}

At early stages of their lives, massive stars are deeply embedded in their native clouds. These stars process their environment with  energetic radiation and stellar winds, creating HII regions and stellar wind driven bubbles \citep{weaver1977}. Overpressure of these regions drives their expansion into the molecular cloud and the HII region changes from hypercompact to ultracompact to compact. Eventually, the ionized gas breaks out, creating a so-called \emph{champagne flow} that disperses the molecular cloud. The interaction between massive stars and their natal cloud controls the star formation process by compressing gas clumps through shocks -- triggering subsequent star formation --, photo-ionizing and photoevaporating envelopes around low mass protostars in the cluster -- creating prominent globules, fingers, and proplyds -- and dispersing gas in ionized and neutral gas flows. This evolution of the region is accompanied by a processing of the cloud material. Specifically, dust coagulates and this has major influence on the absorption properties \citep{ossenkopf1994,ormel2011}.  

The radiation from the powering massive stars is absorbed by the surrounding gas and dust cloud. Only long wavelength (mid-IR to cm) emission from the ionized gas, dust grains, and molecules, and molecular emission can escape and be used to study the properties of the newly formed stars and the early evolution of the region.
In the 3-20 \um~region, strong and broad emission features at 3.3, 6.2, 7.7, 8.6 and 11.3 \um~dominate the spectra \citep{pee02}  due to Polycyclic Aromatic Hydrocarbons (PAHs) molecules \citep{tie08}. These broad emission features are perched on a mid-IR continuum due to transiently heated very small grains (VSG) whose nature is not well understood \citep{desert1990}. At longer wavelengths ($\sim$30~\um), emission by big grains (BG) -- in radiative equilibrium with the radiation field -- take over \citep{desert1990}. Each of these emission components has been ``calibrated'' as a star formation rate indicators on regions of star formation in nearby galaxies \citep{cal07} but these empirical validations have no general theoretical underpinning.


Here, we present mid-IR observations of the deeply embedded OB star cluster in the W3 main region obtained by FORCAST on SOFIA in order to understand the interrelationship of the different mid-IR emission components, their carriers and their evolution in regions of massive star formation. W3 main is a star formation region located towards the Perseus arm at a distance of $1.95\pm 0.04$~kpc, accurately measured using maser kinematics \citep{xu06,hac06}.
Infrared and radio observations have detected a number of bright sources in the W3 main cloud \citet{haw76,wwbn72,tie97}. These objects span a range in evolutionary history with deeply embedded protostars, ultracompact \&\ compact HII regions, all associated with a newly forming cluster containing some 15 OB stars \citep{ojh04,bik11}. A large, embedded population of intermediate- to low-mass stars coexists with the OB stars \citep{megeath1996}. In this paper we focus on W3A, a shell-like compact HII region, powered by IRS2, a 2-3 Myr old O6.5V star \citep{ojh04,bik11} that provides a classic example of a wind driven bubble HII region in an early stage.

\section{Observations and data}
\begin{figure}[t]
\includegraphics[width=\columnwidth]{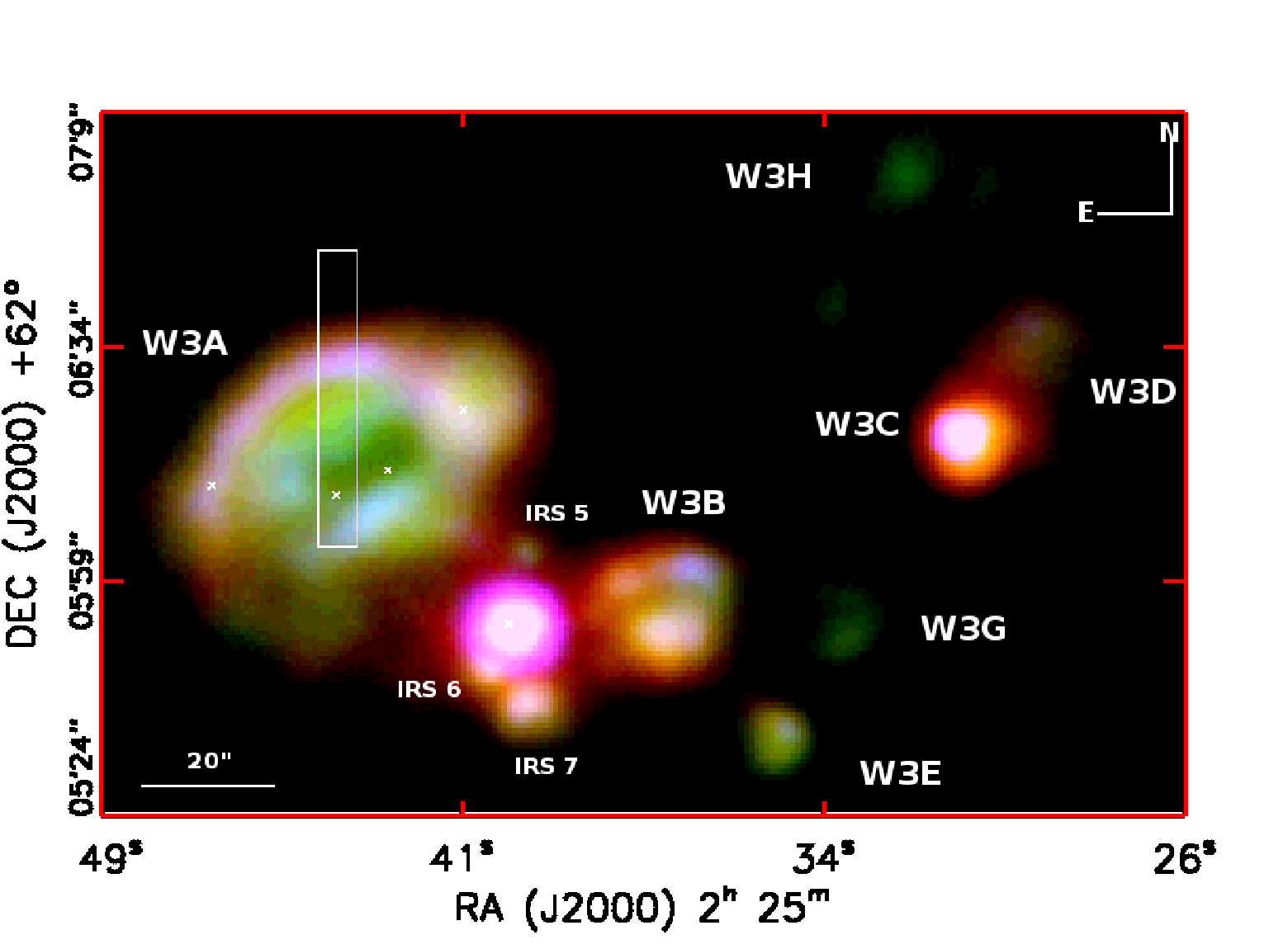}
\caption{Color image of W3 main region, colors are: blue 7.7 \um, green 19.7 \um, and red 37.1 \um. Sources have been identified following \citet{wwbn72} and \citet{tie97}. The location of the four embedded OB stars in W3A are indicated by crosses and are IRS2c, IRS2, IRS2a, and IRS2b (left to right). The rectangular box indicates the direction of the cross cut for which the photometry is shown in Figure~\ref{cross-cuts}.\label{color-image}}
\end{figure}

\begin{figure}[t]
\includegraphics[width=\columnwidth]{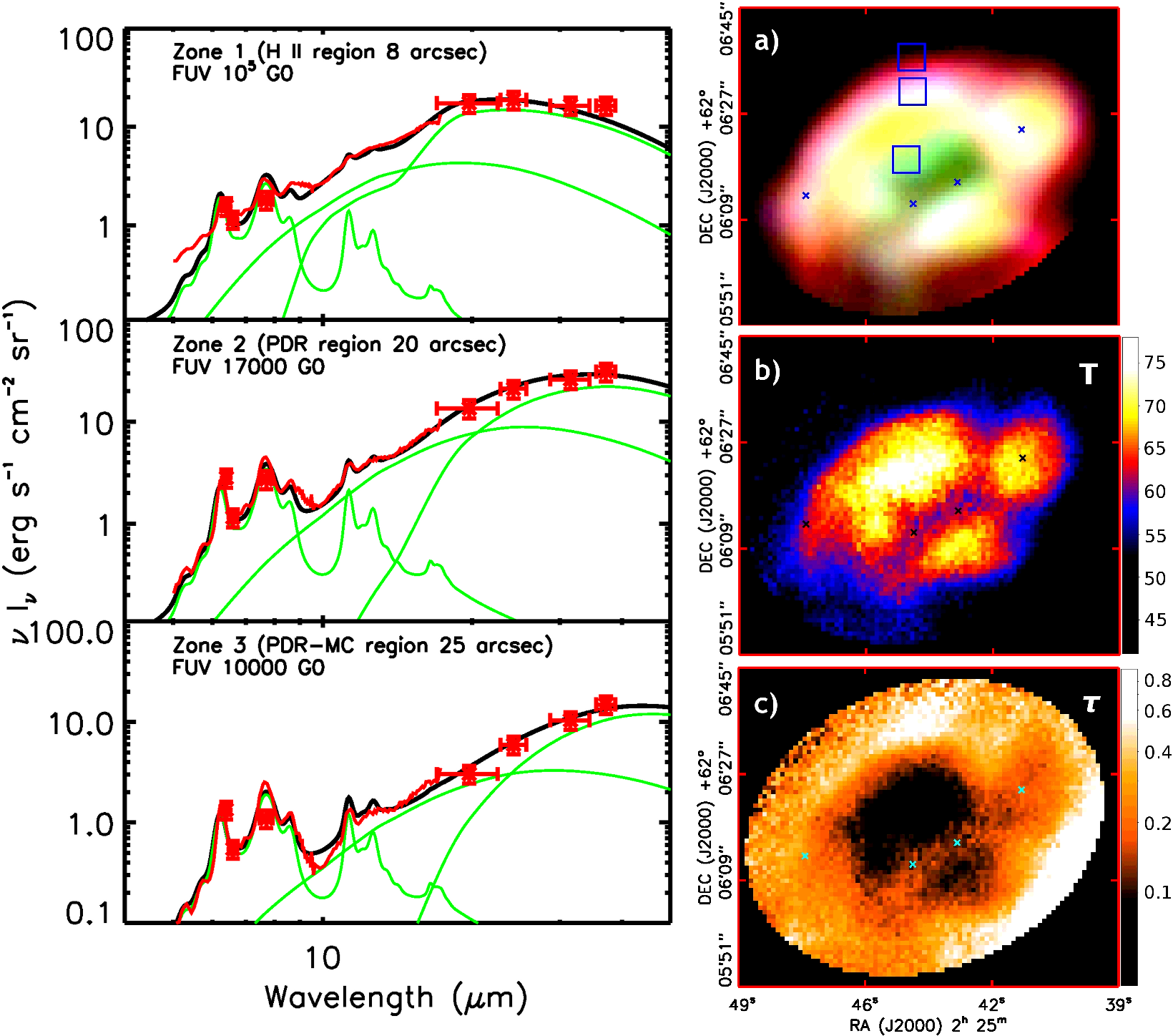}
\caption{\emph{Left}: Spectral Energy Distributions derived from the FORCAST observations (red crosses) and  ISOCAM-CVF spectra (red line) in the three positions marked with squares in 2a. For a clearer comparison, the contributions from the ionized gas emission lines have been subtracted from the CVF spectra. Best fit DUSTEM models are shown as black lines. The green lines show the separate contributions from PAH molecules, Very Small Grains, and Big Grains (peaking left to right). \emph{Right}: a) False color image of the W3A compact HII region. The three squares indicate three positions where spectral energy distributions have been obtained, and crosses mark the position of IRS2c, IRS2, IRS2a and IRS2b from left to right. b) Color temperature map, the peak in emission is located at the north of IRS2 (section 3.1). c) Optical depth map at 37.1 \um.\label{morphology-seds}}
\end{figure}

W3 main was observed by SOFIA during the Short Science flight series on December 4 and  8, 2010, at an altitude of $\sim$43000 feet. We acquired images at 6.4, 6.6, and 7.7 microns in single channel mode and 19.7, 24.2, 31.4, and 37.1 microns using the dichroic beamsplitter in dual channel mode. The field of view of the instrument is the same for all the images, $\sim$3.4\arcmin$\times$3.2\arcmin, and the pixel size is 0.768\arcsec. The point spread function has a FWHM of 2.8\arcsec at wavelengths up to 20 $\mu$m~and is diffraction limited at the longer wavelengths filters \citep{herter2012}. We used a chop throw of $\sim$5\arcmin~and a chop frequency of $\sim$5 Hz. The integration time for each chop pair was $\sim$30 s. The data were processed and calibrated using the pipeline described in \citet{herter2012}. The processed images were aligned and averaged using an iterative sigma rejection algorithm to produce images with effective integration times of 150 s for 6.4 and 6.6 $\mu$m, 270 s for 7.7 $\mu$m, 240 s for 19.7 $\mu$m, 150 s for 24.2 and 31.4 $\mu$m, and 330 s for 37 $\mu$m. The scale size and orientation of the resulting images have been checked against Spitzer IRAC images \citep{ruch2007} and agree well. The estimated $3\sigma$ uncertainty in the calibration due to variations in flat field, water vapor burden, and altitude is approximately $\pm 20\%$. A false color image of the whole W3 main region is shown in Figure~\ref{color-image} and a zoom in on W3A is presented in Figure~\ref{morphology-seds}a.

The FORCAST photometry was checked against the W3-IRS5 spectrum obtained by the Short Wavelength Spectrometer (SWS) on board of the Infrared Space Observatory (ISO)\citep{boogert2008}. W3 IRS5 is a point source in our images and is isolated even within the large SWS beam. At all wavelengths, the two agree to within the absolute calibration error of 20\%. W3A was also observed with SWS/ISO \citep{pee02}. We integrated the emission over the FORCAST filters in the relevant apertures of the SWS spectrum, and found an excellent agreement in flux ($\sim$20\%) at all wavelengths except for the 6.4 \um~band where the FORCAST flux is a factor 1.9 larger. Given that SWS calibration of extended sources is quite challenging, we deem this agreement excellent. Strong [SIII] and [NeIII] lines are present in the SWS spectrum, but the flux contribution of these lines to the FORCAST 19.7, 31.5 and 37.1 bandpasses is less than 6\%. We show cross cuts towards the North in the different FORCAST images in Figure~\ref{cross-cuts}.

W3A was also observed with ISOCAM-CVF. We use the Highly Processed Data Products (HPDP) from the ISO archive \citep{bou05}. Comparison shows that the FORCAST fluxes, \fa~and \fc, are dominated by the PAH 6.2 \um~and 7.7 \um~features and the \fb~traces the underlying plateau \citep{pee02}. By subtracting the \fb~contribution from the \fa~and \fc~bandpasses, we derive the ``pure'' PAH emission.

\section{Results}

\subsection{Observational analysis of W3A}

Figure~\ref{color-image} shows a false color image of W3 main, combining three mid-infrared bands tracing the PAHs and warm dust. All the well known, bright infrared and radio sources are visible in the map, including the deeply embedded protostars associated with W3 IRS5, the ultracompact HII regions W3B, W3C, and W3E, and the compact HII regions, W3A, W3B, W3D, and W3H.  In all, some 10 regions powered by bright O or early B stars are discernable in the mid-infrared. The W3 OB cluster of sources is embedded in low-level PAH emission in the deep 8 \um~IRAC images. However, rather than UV photons leaking out of the confines of the (ultra)compact sources, this emission may be powered by distributed B stars in a more extended cluster, postulated on the basis of the extended [CII] 157 \um~emission \citep{howe}.

\begin{figure}[t]
\includegraphics[angle=90,width=\columnwidth]{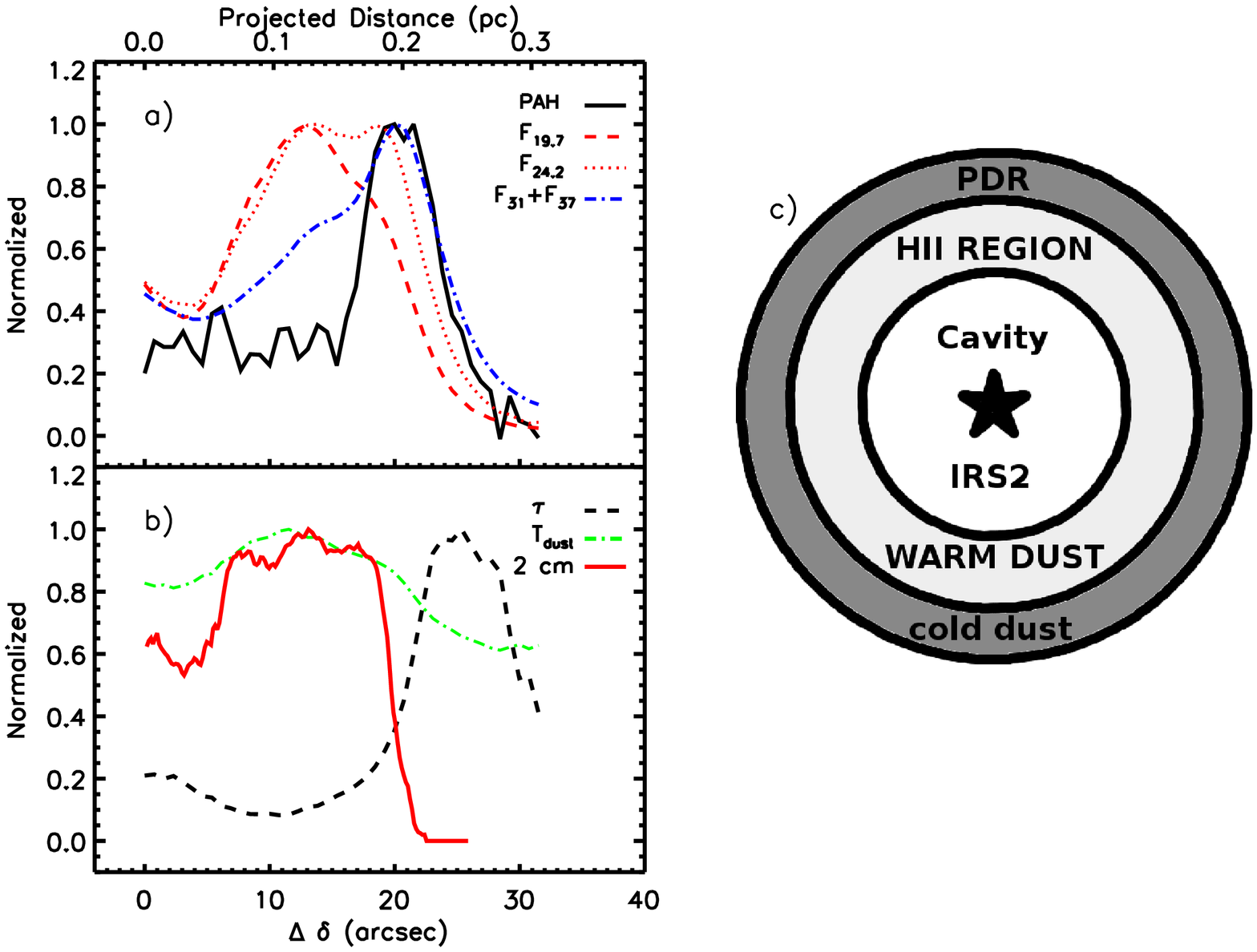}
\caption{a) Cross cuts along the rectangle in Figure~\ref{color-image} as a function of projected distance from IRS2. All values have been normalized to the maximum along the cross cut. $F_{6.4}^{line}+F_{7.7}^{line}$~(black); $F_{19.7}$, $F_{24.2}$~(red); $F_{31.5}+F_{37.1}$~(blue). b) Derived color temperature and optical depth at 37\um~(see text for details), along with the 2 cm emission from \citet{tie97}. c) Schematic plot of W3A, the relative emission from the PDR increases as line of sights are further away from IRS2. \label{cross-cuts}}
\end{figure}

In the radio, the compact HII region, W3A, shows a clear shell-like structure to the north \citep{wwbn72,tie97}. At mid-IR wavelengths, W3A shows a very similar structure (Fig.~\ref{morphology-seds}) and the \fn, \fe~brightness distribution tracks the radio emission well (cf. Figure \ref{cross-cuts}). W3A seem to be a classic example of a wind-driven bubble expanding and sweeping up the surrounding cloud material \citep{weaver1977}.  The radio emission and the \fn~and \fe~warm dust emission originate from the swept up dense ionized gas. This shell is surrounded by a dense PDR (at $\sim$20\arcsec~from IRS 2) traced by the PAH emission (Figure \ref{cross-cuts}). The sharpness and the large relative variations of the PAH emission implies that the PDR is very thin ($\sim$1\arcsec; 0.01 pc) and unresolved. The shell-like structure of this region produces strong limb-brightening and the residual \fn, \fe~emission inside the cavity and the PAH emission seen within the ionized zone reflect this geometry. The inner cavity is presumably filled with hot gas and devoid of dust. However, there is no evidence for X-ray emission from the gas \citep{hof02} possibly due to the effect of mass loading from embedded clumps \citep{redman98}. We deem the alternative model -- ``leakage'' of the hot gas into the surroundings \citep{har09} -- less likely in view of the thin, dense shell enwrapping the ionized shell in the IR images. 

We used the \ff~to \fg~flux ratio to derive color dust temperatures (\td) and optical depths at 37 \um~($\tau_{37}$), assuming a modified black body for the dust ($\kappa_d \sim\nu^{\beta}$~with $\beta=1.8$; \citealt{abe11}). Total dust column density maps have been derived from $\tau_{37}$~using the extinction curve taken from \citet{dra03} with $R_V=4$~(Fig.~\ref{morphology-seds}). Results are summarized in Table 1. For $R_V=3.1$~(5), the column density decreases (increases) by a factor of 0.92 (1.1). Color temperatures and optical depths derived using different filter combinations show substantially the same global structure but the absolute values  increase by 20\% (\fe-\fg) and 30\% (\fn-\fe). We prefer the \ff-\fg~color combination because this emission is expected to originate from the same dust population (i.e. BG; cf., Figure \ref{morphology-seds}). We note that \td~peaks at the ionized gas shell rather than at the IRS2 position, which likely reflects an increase in the relative importance of emission by the cold-dust in the PDR as compared to the warm-dust in the ionized-shell for sightlines traversing the cavity (see section~\ref{section: lyman alpha}). 

The 6.2 and 7.7 \um~PAH emission and the 19-24 \um~emission observed in W3A are associated with the neutral PDR and ionized gas, respectively. It is well established that the PAH emission in compact HII regions peaks in the PDR \citep{tie1993,ber09}. The Spitzer/MIPSGAL \&\ GLIMPSE surveys have shown that many HII regions present bright MIPS 24 \um~emission associated with the ionized gas and are surrounded by a shell of bright 8 \um~PAH emission \citep{anderson2011,carey2009}. With this in mind, we recognize that the star W3 IRS2b is also surrounded by a shell of PAH emission (Figure~\ref{morphology-seds}) while  bright 19 \&\ 24 \um~emission is present in the ionized gas. We thus surmise that this star is still embedded in its own HII region, separate from W3A. Perusal of JHK images of W3A \citep{bik11} also suggests that the emission from IRS2b is separate from that of the main HII region. IRS2 has already ``swallowed'' the HII region associated with IRS2a and IRS2c and likely the region around IRS2b is next on the menu. We consider that the ridge of bright emission towards the south of IRS2a is still a remnant of the earlier, isolated HII regions phase of this star. 

From the FORCAST data, the region associated with IRS2b (O8V) has a $6-40$~\um~luminosity of $1\times10^5~\mathrm{L_{\sun}}$~in good agreement with the estimate based on near-IR photometry and spectroscopy \citep{bik11}. The whole W3A region has a $6-40$~\um~ luminosity of $4.1\times10^5~\mathrm{L_{\sun}}$. Subtracting IRS2b, the observed $6-40$~\um~luminosity of W3A is $\sim$$3\times10^5~\mathrm{L_{\sun}}$. Based on the spectral type of \citet{bik11}, and using \citet{mar05}, we estimate that IRS2 (O6.5V), IRS2a (O8.5V), and IRS2c (B0.5V) emit $1.6\times10^5~\mathrm{L_{\sun}}$, $6.6\times10^4~\mathrm{L_{\sun}}$,~and $4\times10^4~\mathrm{L_{\sun}}$, respectively. The total luminosity of the stars associated with W3A is thus $\sim$$3\times10^5~\mathrm{L_{\sun}}$. We note that the FORCAST-derived luminosity is just a lower limit to the total infrared luminosity, because far-IR photometry is lacking. Nevertheless, it seems that the luminosity of the known OB stars is sufficient to explain the observations and no major contribution from embedded YSOs is required, consistent with \citet{ojh04}. The $6-40$~\um~luminosity associated with the ionized gas shell is $1.1\times10^5~\mathrm{L_{\sun}}$. Hence, we estimate that the (radial) UV (absorption) optical depth of the dust in the ionized gas is, $\tau_{UV}~=-~\ln\left(L_{6-40}/L_{\star}\right)\simeq 1$.

\subsection{Dust Spectral Energy Distribution}

To study the dust characteristics in W3A, three zones were selected to extract Spectral Energy Distributions (SED) in 6\arcsec$\times$6\arcsec~apertures: At 8\arcsec~north of IRS2 inside the HII region; at 20\arcsec~north of IRS2 in the PDR, and in the molecular cloud at 25\arcsec~north of IRS2 (Figure~\ref{morphology-seds}). These SEDs show a pronounced shift of the peak emission towards longer wavelengths when moving from the HII region to the PDR and on to the molecular cloud. The DUSTEM model (version 3.8; \citealt{com11}) was used to fit these SEDs, adopting the grain size distribution for the diffuse high galactic latitude model. 
We adopt an O6V the closest to the spectral type of the central star 
and used the appropriate solar metallicity model \citep{kur93} scaled to match the number of ionizing photons \citep{tie97}, modified to our adopted distance of 1.95 kpc. We have evaluated these stellar fluxes at the projected distances of the zones. For the HII region position, we included the FUV and EUV stellar emission, while for the PDR and molecular cloud zones only FUV is included. In addition, the molecular cloud radiation field has been attenuated by a radial optical depth of $\tau_{UV}=1$. Finally, IR fluxes for each of the DUSTEM model components (PAH, VSG and BG) are adjusted to fit the observations. Resulting fits are shown in Figure~\ref{morphology-seds} and derived parameters are summarized in Table 1. The results illustrate that the 31.5 and 37.1 \um~emission originates from BG, which also dominate the 24 \um~emission from the ionized shell. The 24 \um~emission in the PDR has an important contribution from VSGs.
The good agreement between the model results and observation for the long wavelength emission (i.e. BG), demonstrates that the adopted radiation fields -- the main free parameters -- are reasonable.
\begin{deluxetable}{lccc}
\tablewidth{\columnwidth}
\tablehead{
\colhead{Parameters} & \colhead{H II} & \colhead{PDR}   & \colhead{Mol. Cloud}\\
                     & (0.08 pc)      & (0.2 pc)        & (0.25 pc)      
}
\startdata
$F_{\ast}~(ergs~s^{-1}~cm^{-2})$               & 247   & 27  &  7       \\
\cutinhead{Observational results}
$\tau_{37.1}$                                 & 0.06  & 0.24  & 0.55   \\
$N_\mathrm{Dust}~ (10^{-6} g~cm^{-2})$          & 211   & 830   & 1921   \\
$N_H~ (10^{21}~cm^{-2})$                       & 7.8  & ...   & ...    \\
$T_\mathrm{color}~ (K)$                        & 72    & 64    & 50   \\
Dust-to-gas mass ratio                       &0.012  & ...   & ...  \\
\cutinhead{DUSTEM results}
$T_\mathrm{Dust}(\mathrm{100~nm})~(K)$         & 78    & 55     & 44  \\
L$_{PAH}$/L$_{IR}~(10^{-2})$                    & 0.97  & 3.74  & ...  \\
$A_V$                                        & 2.1   & 32     & 61  \\
R$_V$                                        & 4.2   & 4.0    & 4.3 
\enddata
\end{deluxetable}

\subsection{Lyman alpha dust heating}
\label{section: lyman alpha}

The dust temperature peaks on the ionized gas shell and this is very reminiscent of Ly$\alpha$ heating. From the radio observations \citep{tie97}, we derive the Ly$\alpha$~luminosity of $3.2\times10^4\mathrm{L_{\sun}}$~or only 30\% of the observed $6-40$~\um~luminosity associated with the ionized gas shell. In general, the contribution of trapped Ly$\alpha$~photons to the dust heating ($\Gamma_\alpha$) relative to the stellar radiation ($\Gamma_\star$), is approximately given by $\Gamma_\alpha/\Gamma_\star = (r/R_S)^2\times 1/\tau_{UV}$~\citep{tie05} with $r$~the distance from the star (0.12 pc), $R_S$~the Str\"omgren radius of the region (0.2 pc) and $\tau_{UV}\sim$1~the radial optical depth of the dust (see above). In the middle of the ionized gas shell,  Ly$\alpha$~heating contributes $\sim$25\% of the dust heating.

We used Cloudy version 10.00 \citep{fer98} to model the W3A \td~distribution and study the heating due to trapped Ly$\alpha$~photons. Assuming an O6 star with a luminosity of $2\times10^5 \mathrm{L_{\sun}}$, and a shell density of 5000 cm$^{-3}$, the model shows that only for $\tau_{UV}\lesssim$0.2 (a factor of 5 less than observed) dust heating is dominated by Ly$\alpha$~and the temperature distribution tracks the density structure. Hence, both the observed luminosity and the observed optical depth imply that stellar radiation dominates the temperature distribution. We attribute the rather constant dust temperature in the ionized shell to the morphology of the region. Specifically, the absence of a rise of the dust temperature towards the central star reflects the presence of an inner dust-free cavity. The geometry of a dust-free cavity, surrounded by a dusty ionized gas shell and PDR (Figure \ref{cross-cuts}b), will tend to give a rather constant color temperature with projected distance from the star, because the observed SED is a superposition of all emission components (ionized shell \&\ PDR) along the line of sight.

\subsection{Dust-to-gas ratio}

The H-nuclei column density was calculated from the radio \citep{tie97}, assuming a shell-like geometry, to be $N_H=7.8\times10^{21}$ cm$^{-2}$. The dust column density derived with DUSTEM results then in a dust-to-gas mass ratio of $0.012$, close to the diffuse ISM value. In contrast, the radial UV optical depth derived from the
observed $6-40$~\um~luminosity ($\tau_{UV}$=1, section 3.1) translates into a dust-to-gas mass ratio of $0.035$~using the \citet{dra03} $R_V=4.0$~model.
While this difference seems small, we consider it significant. To phrase this differently, adopting dust properties appropriate for the diffuse ISM, the radial absorption optical depth by dust in the ionized gas would be $\tau_{UV}\simeq$4~and only $\simeq$2\%~of the number of ionizing photons emitted by the central star would be used to ionize the gas \citep{petrosian}. This would imply an enormous ionizing photon luminosity that is incompatible with the observed properties of the ionizing stars in this HII region and the observed IR luminosity of the dust. This is a well-known issue, dating back to the earliest IR studies of HII regions \citep{tielens1979}, and it seems that the dust properties in the ionized gas have to be different from those in the diffuse ISM. This may reflect either the selective destruction of PAHs and VSGs in the ionized gas -which are important contributors to the FUV dust extinction- or more generally an increase in grain size during the preceding molecular cloud core phase driven by coagulation \citep{ossenkopf1994,pagani2010,ormel2011}.

\acknowledgments

This work is based on observations made with the NASA/DLR  
Stratospheric Observatory for Infrared Astronomy (SOFIA). SOFIA  
science mission operations are conducted jointly by the Universities  
Space Research Association, Inc. (USRA), under NASA contract  
NAS2-97001, and the Deutsches SOFIA Institut (DSI) under DLR  
contract 50 OK 0901. Financial support for FORCAST was provided to  
Cornell by NASA through award 8500-98-014 issued by USRA.
Studies of interstellar PAHs at Leiden Observatory are supported through advanced-ERC grant 246976 from the European Research Council and through the Dutch Astrochemistry Network funded by the Dutch Science Organization, NWO.

\end{document}